\documentclass[aps,pre,twocolumn, superscriptaddress,showpacs,floatfix]{revtex4}
\usepackage{amsmath,amssymb,graphicx,color}
\usepackage[flushleft]{threeparttable}
\usepackage[normalem]{ulem}
\begin{document}
\title{Fluctuations of red blood cell membranes: The role of cytoskeleton}

\author{Wonjune Choi}
\affiliation{Department of Physics, Korea Advanced Institute of Science and Technology, Deajeon 305-701, Korea}

\author{Juyeon Yi}
\affiliation{Department of Physics, Pusan National University, Busan 609-735, Korea}

\author{Yong Woon Kim\footnote{corresponding author: y.w.kim@kaist.ac.kr}}\affiliation{Graduate School of Nanoscience and Technology, Korea Advanced Institute of Science and Technology, Deajeon 305-701, Korea}

\begin{abstract}
We theoretically investigate the membrane fluctuations of red blood cells with focus laid on the role of the cytoskeleton, viewing the system as a membrane coupled to sparse spring network. This model is exactly solvable and enables us to examine the coupling strength dependence of the membrane undulation. We find that the coupling modifies the fluctuation spectrum at wavelengths longer than the mesh size of the network, while leaving the fluid-like behavior of the membrane intact at shorter wavelengths. The fluctuation spectra can be markedly different, depending on not only the relative amplitude of the bilayer bending energy with respect to the cytoskeleton deformation energy but also the bilayer-cytoskelton coupling strength. 
\end{abstract}


\maketitle

\section{Introduction}

A red blood cell~(RBC) has been a standard test bed of membrane physics due to its simple structure lacking a nucleus and organelles. Early studies, treating a RBC as a lipid bilayer bag, already rendered insightful understanding of various properties, to name a few, biconcave shapes\cite{Canham, Deuling}, flickering phenomenon~\cite{Brochard}, and tank-treading motion~\cite{Keller}. Still there are certain features signifying the role of other substructure; a spectrin network (cytoskeleton) coupled to the bilayer is found to be responsible for the shape transformation sequence~\cite{RBC shape}, large-scale shape changes under shear flow~\cite{Peterson}, and crenated shapes~\cite{Echinocyte}. Many attempts have been made in order to provide simplified descriptions of the cytokeleton as for its effects on non-trivial shapes and fluctuation of RBCs~\cite{Park model, Nelson, Li, Rochal, Auth}. 

Meanwhile, theoretical works~\cite{Gov 2003,Fournier} analyzed the fluctuation spectrum measured by Zilker {\it et al.}~\cite{Zilker} and introduced a picture of a composite membrane, that is, a lipid bilayer sparsely connected to the cytoskeleton.
In Ref.~\cite{Gov 2003}, through an empirical approach, it was claimed that the coupling to the cytoskeleton induces surface tension in such a way that the effective bending modulus of the lipid bilayer undergoes an abrupt jump at the crossover length~\cite{Gov 2003}.
This claim was elaborated by considering the cytoskeleton explicitly as spring meshwork~\cite{Fournier}, where the authors examine the
elastic energy of the meshwork as a function of the membrane area coarsely grained at the mesh size $\xi$ and thereby conclude that the tension contribution 
vanishes suddenly at length scale $\xi$.  Also, a model Hamiltonian was proposed, assuming infinitely strong coupling of
the membrane at sites linked to the cytoskeleton, so as to explain the induced tension in the long wavelength fluctuations~\cite{dubus}.   

In this regard it seems that a RBC as a composite membrane necessarily acquires surface tension at long length scales. 
However,  clear signature of the surface tension was not always observable, and also there exists subtle variance between recent observations. Yoon {\it et al.}~\cite{Yoon} reported the shape dependence of the fluctuation spectrum where the tension contribution is fairly pronounced for spherocype-shaped RBCs  but almost unnoticeable for discocytes. 
In contrast, Popescu {\it et al.} observed clearly the induced surface tension irrespective of shapes~\cite{Popescu}.  Using similar interferometric technique in Ref.~\cite{Popescu}, many red blood cells showed bending dominant fluctuations at short wavelengths and confining dominant long wavelength fluctuations, but the intermediate tension dominant region could be seen from few number of the samples \cite{private}. In all those experiments, the notion of tension jump seems to be not evident as derived from the theories~\cite{Gov 2003, Fournier}. The origin of such diversity is left unknown, and more experimental evidences together with theoretical models are anticipated in order to draw a unified picture of a RBC as a composite membrane.

In this study, we propose an exactly solvable model to describe elastic properties of a RBC membrane, positing that the cytoskeleton is spring network and the membrane is sparsely and regularly linked to the network via harmonic coupling. 
Unlike assumed in Ref.~\cite{dubus}, the coupling strength is finite and may even be weak due to flexibility of protein complex linking the bilayer and the cytoskeleton and also due to fast dissociation/association process of the cytoskeleton controlled by ATP concentration \cite{Manno, Betz}.  Although governing factors are not known precisely, it is reasonable to think that
the coupling strength can vary depending on experimental conditions, and observation results should intrinsically be diverse.  In our consideration, the coupling strength is
introduced as an essential model parameter which interplays with other involved energy scales to determine the tension emergence. 

This paper is organized as follows. In Sec. II, we introduce a model describing energies associated with fluctuations of the membrane, deformation of the spring network, 
and the coupling between them. The effective Hamiltonian for the equilibrium fluctuation of the membrane is obtained in Sec. III, and analytic expression of the fluctuation spectrum is derived in Sec. IV. Section V consists of three subsections. In Sec. V. A, various kinds of the qualitatively distinctive fluctuation spectra are analyzed. We find that at length scale shorter than the network mesh size, the cytoskeleton elasticity is negligible irrespective of the coupling strength, and the fluctuation spectrum of the membrane exhibits the fluid-like behavior. At long wavelengths, the coupling to the cytoskeleton imparts elasticity to the membrane, and yet resulting features in the fluctuation spectrum thereof are strongly dependent on the system parameters.
The effect of confining potential is discussed in Sec. V. B. In the last subsection V. C, we compare our results with previous studies, in particular related to the tension ``jump" stated in Ref.~\cite{Fournier, Gov 2003}, and conclusion follows in Sec. VI.

\section{System}
We consider a lipid bilayer of a RBC as an almost flat, symmetric(zero spontaneous curvature) fluid membrane.  Assuming small deformations without overhangs, the membrane height can be represented by a single-valued function $h({\mathbf r})$ where ${\mathbf r}$ spans the flat reference plane at $h({\mathbf r}) =0$. The corresponding Hamiltonian reads as:
\begin{equation}\label{H_m}
{\cal H}_{m}[h]=\int d{\mathbf r} ~\kappa (\nabla^2 h)^2 +\gamma h^{2} ~.
\end{equation}
The second term, so called the confining potential, is introduced phenomenologically to describe the long wavelength fluctuations of red blood cell membranes \cite{Gov 2003}. Closed geometry \cite{Auth, Park model, Nelson}, periodic pinning \cite{Gov 2004} and excluded-volume effect \cite{Gov 2003, Farago} were suggested as the physical origin of the confinement, yet no consensus has been met. Although the term has been widely used for many experimental studies \cite{Betz, Popescu}, there are also skeptical views on the necessity of the confining potential \cite{Yoon}. To embrace every possible situations, we do not discard this phenomenological term from the model. However, overall discussion in this paper will be made with very weak confinement so that the effect of the confinement is marginal to the fluctuation spectrum; the limit $\gamma \rightarrow 0^{+}$ would not hurt the key lessons of our research. More subtle issues on the confining potential would be discussed in Sec. V. B.
We also mention that in order to reflect area conserving property of a bilayer, the bending energy usually includes effective bare surface tension, which is dependent on the excess area of the bilayer and ambient temperature \cite{Milner, Seifert}. However, we intentionally discard the bare tension in the Hamiltonian ${\cal H}_{m}$ in order to explicate the induced tension by the cytoskeleton.

Another ingredient comprising the system is the cytoskeleton which is  two-dimensional regular network of spectrin tetramers. We consider the cytoskeleton as  spring network~\cite{springnetwork} and model it as a sparse square array of beads connected by springs. Since we assumed the small deformations, the in-plane stretching and the out-of-plane deformation is approximately decoupled \cite{dubus}. In addition, the single particle tracking experiment also showed that the diffusion of the band-3 proteins which are connected to the spectrins are significantly restricted \cite{Tomishige}. Hence, we assume that the in-plane position of the $i$th bead is fixed to form a square lattice with uniform spacing $\xi$ and represented by the position vector ${\bf x}_{i} = \xi(m {\hat x}+n{\hat y})$ with integers $m$ and $n$. On the other hand, we allow out-of-plane displacements of the beads and express the displacement of the $i$th bead by $\phi({\mathbf x}_{i})$.
The imbalance in the displacements between the nearest neighboring beads brings about either contraction or dilation of the connecting springs. Assuming the identical spring constant $k$, the elastic energy cost for such deformation of the spring network is,
\begin{equation}\label{H_e}
{\cal H}_{e}[\phi]=\sum_{\langle {\bf x}_{i},{\bf u}\rangle}k[\phi({\mathbf x}_{i})-\phi({\mathbf x}_{i}+{\bf u})]^2~,
\end{equation}
where ${\bf u}=\xi{\hat x}$ or ${\bf u}=\xi{\hat y}$ are the primitive vectors for the square lattice. In fact, the cytoskeleton is closer to triangular (or hexagonal) network. Yet we expect that effects of the symmetry details must be weak on the qualitative properties of the membrane fluctuation and hence pursue
simple analytic approach, remaining in the picture of the four-fold network.

We finally introduce interaction between the membrane and the spring network. The coupling between the two systems is mediated by protein inclusions in the membrane and adaptor proteins~(for example, Ankyrin) bridging the membrane inclusion and the spectrin network. Hence, the coupling must be the consequence of various kinds of microscopic interactions among the membrane, inclusions, adaptor proteins, and spectrins. Rather than deriving the coupling energy from all these complex origins, here we approximate it under the assumption that the separation between the membrane and the coupled cytoskeleton $|h({\mathbf x}_{i})-\phi({\mathbf x}_{i})|$ is small. As the leading order of Taylor expansion, we can then model the coupling energy by harmonic potentials between the out-of-plane displacement of the bead and the height undulation of the membrane at
the bead location ${\bf x}_{i}$:
\begin{equation}\label{V}
V[h,\phi]=\sum_{{\mathbf x}_{i}}v[h({\mathbf x}_{i})-\phi({\mathbf x}_{i})]^2~,
\end{equation}
where $v$ measures the coupling strength.
Since recent numerical research based on the harmonic membrane-cytoskeleton interaction showed good agreement with experiments on red blood cell membranes \cite{Peng}, we expect the harmonic approximation is sufficient and ignore the higher order terms \cite{note}.

In total, the system Hamiltonian of our interest is composed as
${\cal H}= {\cal H}_{m}+{\cal H}_{e}+V$. In Ref.~\cite{dubus}, the authors considered the bilayer-cytoskeleton interaction by imposing a hard constraint that
the membrane height at the coupling site is identical with the bead location, i.e., $h({\mathbf x}_{i}) = \phi({\mathbf x}_{i})$. In our model, this corresponds to an extreme situation of infinite coupling ($v \rightarrow \infty$), and the behaviors at general coupling strengths are still to be investigated. Even though steric interaction can also be involved to correlate the membrane and the cytoskeleton \cite{Auth}, this research neglects the excluded-volume effect and assumes the ``phantom" membrane and network. 

\section{effective hamiltonian for membrane fluctuation}

Equilibrium properties of the system can be extracted from evaluating the canonical partition function:
\begin{eqnarray}\label{Z}
Z&=&\int {\cal D}\phi {\cal D}h ~e^{-\beta {\cal H}[h,\phi]}~ \\ \nonumber
&=& \int {\cal D}h ~e^{-\beta {\cal H}_{\mathrm{eff}}[h]} ~,
\end{eqnarray}
where the second line defines the effective Hamiltonian for the height deformation of the membrane, ${\cal H}_{\mathrm{eff}}[h]$, which
can be obtained by integrating the Boltzmann factor over the fields $\phi({\bf x})$.

For mathematical convenience, let us obtain the Hamiltonian in the momentum space. For the height field $h({\bf r})$ of the membrane, we employ the Fourier transform in two dimensions,
\begin{equation}\label{ft1}
h({\bf x})=\sum_{{\bf q}}e^{i{\bf q}\cdot{\bf x}} h_{{\bf q}}/N ~.
\end{equation}
Here we discretize the space in units of the length $b$ which is a microscopic length scale, e.g., molecular size of a lipid molecule consisting the membrane, and let the momentum ${\bf q}$ take values to satisfy the periodic boundary conditions, $e^{iq_{\alpha}bN}=1$ for $\alpha=x,y$ with $Nb$ being the lateral dimension of the reference plane. We then obtain the bending energy of the membrane in terms of $h_{{\bf q}}$ as
\begin{equation}\label{ft1 H_m}
{\cal H}_{m}[h_{{\bf q}}]=\sum_{{\bf q}}E_{m}({\bf q}) |h_{{\bf q}}|^{2}~,
\end{equation}
where the bending energy spectrum in the discrete space is given by $E_{m}({\bf q})=(4\kappa/b^{2})[2-\sum_{\alpha=x,y}\cos(q_{\alpha}b)]^{2}+\gamma$.
Since we are interested in the length scale much longer than the size of the monomer, we take the continuum limit and describe the bending energy of
the bilayer as $E_{m}({\bf q})=\kappa|{\bf q}|^{4}+\gamma$.

The remaining components of the Hamiltonian are given in terms of the fields defined only at the link sites ${\bf x}_{i}$'s,
for which we introduce the Fourier transform as
\begin{equation}\label{ft2}
f({\bf x}_{i})= \sum_{{\bf q}}e^{i{\bf q}\cdot{\bf x}_{i}} {\widetilde f}_{{\bf q}}/M,~~~ f= h, \phi~.
\end{equation}
This is again the discrete Fourier transform but the length unit is given by $\xi$ instead of $b$. In order to discern the Fourier transform,
Eq.~(\ref{ft1}), we give the tilde symbol to the function in the momentum space.
Here a commensurate condition $\xi = pb$ with an integer $p$ is assumed, and $M=N/p$ is the number of link sites along one direction. For the periodic boundary conditions, the values of $q_{\alpha}$ are given by integer multiples of $2\pi/L$ but lie in the restricted Brillouin zone~(RBZ),
\begin{equation}\label{rbz}
-\pi/\xi < q_{\alpha} < \pi /\xi~,
\end{equation}
where the zone boundary $q_c = \pi/\xi$ reflects the discrete nature of the spring network in our theory.

With Eq.~(\ref{ft2}) applied, the elastic energy of the spring network, Eq.~(\ref{H_e}), is diagonalized into
\begin{equation}\label{ft2 H_e}
{\cal H}_{e}[{\widetilde \phi}_{{\bf q}}]=\sum_{{\bf q}\in \mathrm{RBZ}}E_{e}({\bf q})|{\widetilde \phi}_{{\bf q}}|^{2}~,
\end{equation}
where $\sum_{{\bf q}\in \mathrm{RBZ}}$ denotes the summation over the range given in Eq.~(\ref{rbz}), and the energy spectrum is given by $E_{e}({\bf q})=\sum_{\alpha=x,y}2k[1-\cos(q_{\alpha}\xi)]$, which gives $|{\bf q}|^{2}$ behavior only in the long-wavelength limit specified
by the length scale $\xi$: $q_{\alpha}\xi \ll 1$. On the other hand, the coupling energy, Eq.~(\ref{V}), is transformed as
\begin{equation}\label{ft2 V}
V[{\widetilde \phi}_{{\bf q}},{\widetilde h}_{{\bf q}}]=
\sum_{{\bf q}\in \mathrm{RBZ}} v|{\widetilde \phi}_{{\bf q}}-{\widetilde h}_{{\bf q}}|^2~.
\end{equation}

Given the Hamiltonian components in the momentum space by Eqs.~(\ref{ft1 H_m}),~(\ref{ft2 H_e}), and (\ref{ft2 V}), we perform integration in Eq.~(\ref{Z}) over the field ${\widetilde \phi}_{{\bf q}}$ and obtain the effective Hamiltonian governing the membrane height fluctuations as follows:
\begin{equation}\label{H_eff1}
{\cal H}_{\mathrm{eff}}[h]=\sum_{{\bf q}}E_{m}({\bf q})|h_{{\bf q}}|^{2}+\sum_{{\bf q}\in \mathrm{RBZ}}
\frac{vE_{e}({\bf q})}{v+E_{e}({\bf q})}|{\widetilde h}_{{\bf q}}|^{2}~.
\end{equation}
The second term induced by the coupling to the spring network describes the energy for the height deformations
occurring over length scales longer than $\xi$. In the infinite coupling limit, one can find that the coefficient of
$|{\widetilde h}_{{\bf q}}|^{2}$ becomes $E_{e}({\bf q})$, which yields harmonic couplings in the real space
representation, $[h({\bf x}_{i})-h({\bf x}_{i}+{\bf u})]^2$, as considered in Ref.~\cite{dubus}. For finite coupling strength, expanding the coefficient as a series of
 $1/v$ gives the first-order term, $-v^{-1}E^{2}_{e}({\bf q})$, and the corresponding energy is given by a lattice Laplacian, $-[h({\bf x}_{i}+{\bf u})+h({\bf x}_{i}-{\bf u})-2h({\bf x}_{i})]^2$, where the negative sign indicates energy gain to create large-scale ($\xi$) curvature.

\section{fluctuation spectrum}

Now we investigate the fluctuation spectrum of the membrane in equilibrium, $\langle |h_{\bf q}|^{2}\rangle$. The angular bracket denotes the equilibrium average defined as
\begin{equation}\label{define psd}
\langle |h_{\bf q}|^{2}\rangle = Z^{-1}\int {\cal D}h_{{\bf q}} ~e^{-\beta {\cal H}_{\mathrm{eff}}[h_{{\bf q}}]}h_{{\bf q}}h_{{\bf q}}^{*}~,
\end{equation}
where the partition function $Z$ and the effective Hamiltonian ${\cal H}_{\mathrm{eff}}[h_{{\bf q}}]$ are given in Eqs.~(\ref{Z}) and (\ref{H_eff1}), respectively. In order to perform the integration, we need to express  ${\widetilde h}_{{\bf q}}$ in the effective Hamiltonian~(\ref{H_eff1}) in terms of $h_{{\bf q}}$. Reciprocal lattice vectors ${\bf T}=(2\pi/\xi) (n_{x}{\hat x}+n_{y}{\hat y})$ with integral coefficients $n_{x}$ and $n_{y}$ are defined by the orthogonality relation for the discrete lattice $\{ \mathbf{x_i} \}$:
\begin{equation}\label{ortho}
\sum_{{\bf x}_{i}}e^{i\mathbf{q}\cdot {\mathbf x}_{i}} = (N/p)^2\delta_{{\bf q},{\bf T}}~,
\end{equation}
which gives the inverse transform of Eq.~(\ref{ft2}),
\begin{equation}
{\widetilde h}_{\bf q} = (M)^{-1}\sum_{{\bf x}_{i}}e^{-i\mathbf {q}\cdot {\mathbf x}_{i}}h({\bf x}_{i})~.
\end{equation}
Here we can replace $h({\bf x}_{i})$ with the right hand side of Eq.~(\ref{ft1}) with ${\bf r}={\bf x}_{i}$. Then, from Eq.~(\ref{ortho}) it follows that
\begin{equation}
{\widetilde h}_{\bf q}=p^{-1}\sum_{{\mathbf T}}h_{{\mathbf q}+{\mathbf T}}~,
\end{equation}
where ${\bf q}$ is restricted to the range in Eq.~(\ref{rbz}).
Using this relation, we write the effective Hamiltonian:
\begin{equation}\label{H_eff2}
{\cal H}_{\mathrm{eff}}[h_{{\bf q}}]=\sum_{{\bf q}\in \mathrm{RBZ}}\sum_{{\bf T},{\bf T}'}
h^{*}_{{\bf q}+{\bf T}}M_{{\bf q}}({\bf T},{\bf T}')h_{{\bf q}+{\bf T}'}~,
\end{equation}
where the summation over ${\bf q}$ in the bending energy term is folded into the restricted Brillouin zone (RBZ), and the interaction kernel is given by
\begin{equation}\label{H_eff3}
M_{{\bf q}}({\bf T},{\bf T}')=E_{m}({\bf q}+{\bf T})\delta_{{\bf T},{\bf T}'}+C({\bf q})
\end{equation}
with
\begin{equation}
C({\bf q})=\frac{1}{p^{2}}\frac{vE_{e}({\bf q})}{v+E_{e}({\bf q})}~.
\label{C_q}
\end{equation}
Note that $C({\mathbf q})$ induced by the bilayer-network coupling describes the all-to-all coupling between any deformation fields with their wave vector difference given by the reciprocal lattice vector of the spring network, reflecting the fact that length scales shorter than $\xi$ are indiscernible as far as the network effect is concerned.

With the effective Hamiltonian given by Eq.~(\ref{H_eff2}), the integration in Eq.~(\ref{define psd}) is performed to give 
\begin{align}\label{psd}
\langle |h_{{\bf q}}|^{2}\rangle 
&=\frac{k_\mathrm{B} T}{E_{m}({\bf q})}\left\{1-\frac{C({\bf q})E^{-1}_{m}({\bf q})}
{1+C({\bf q})\sum_{{\bf T}}E^{-1}_{m}({\bf q}+{\bf T})}\right \}  \nonumber \\
&=\frac{k_\mathrm{B} T}{E_{m}({\bf q})+C_{\mathrm{eff}}({\bf q})}~,
\end{align}
with
\begin{equation}
C^{-1}_{\mathrm{eff}}({\bf q}) = C^{-1}({\bf q})+\sum_{{\bf T}\neq {\mathbf 0}}E^{-1}_{m}({\bf q}+{\bf T})~,
\label{C_eff}
\end{equation}	
which is the main results of our model for red blood cell membranes. Two factors enter the characteristics of the fluctuation spectra: The all-to-all coupling energy kernel $C(\mathbf{q})$ and the bending energy kernel for wave vectors in the high-order Brillouin zone $E_m (\mathbf{q+T})$. Because the effective cytoskeleton elastic energy $C_{\mathrm{eff}}(\mathbf{q})$ is the harmonic mean of these two terms, the smaller energy contributes more to the fluctuation spectrum. This competition of two factors can result in new kinds of fluctuation spectra of the coupled membrane, which will be discussed in the next section.

\section{Results and Discussion}
\subsection{Diversity of Fluctuation Spectrum}

Although the phenomenological model proposed by Gov \textit{et al}. explains the long wavelength fluctuation data of Ref.~\cite{Zilker}, the bilayer coupled to the two-dimensional meshwork may not follow such description for all possible set of parameters $\kappa$, $k$, and $v$, and it may possess potential diversity in its fluctuation spectrum.  As given in Eq.~(\ref{psd}), the characteristics of the fluctuation spectrum are determined by 
three competing energies: $E_{m}({\mathbf  q})$, $E_{m}(\mathbf {q +T})$, and $C({\mathbf q})$. Roughly speaking, for given $q$ and the system parameters,
one of the energy functions, which satisfies 
\begin{equation}\label{cri}
\mbox{max}[ E_{m}({\mathbf q}), \mbox{min}[C(\mathbf{q}), E_{m}(\mathbf{q+T})]~,  
\end{equation}
determines the main properties of the fluctuation spectrum. Here the coupling strength $v$ regulates the form of 
$C({\mathbf q})$.  
It should be also mentioned that the fluctuation spectrum has directional dependence which can be removed by angular average. We have confirmed that the average does not show any qualitatively different properties apart from a constant factor. Every different type of fluctuations (e.g., non-monotonic fluctuation spectrum) can also be observed from the angular averaged fluctuation spectra. We will thus let ${\mathbf q}=(q,0)$ henceforth.  
\begin{figure}[!thb]
\resizebox{8.5cm}{!}{\includegraphics{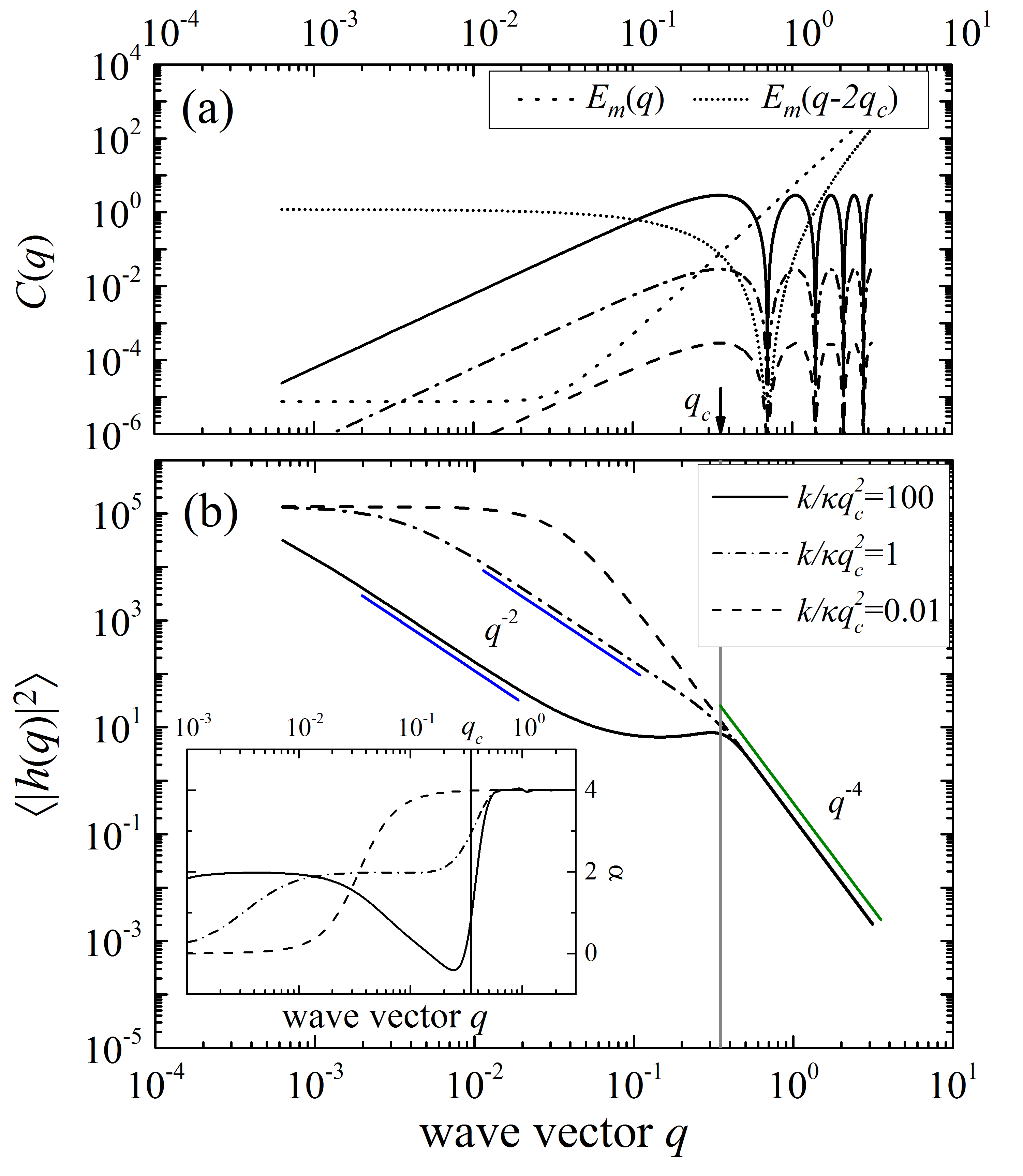}}
\caption{Relevant energy functions and fluctuation spectra in the strong coupling case~($v=100k$), where the system parameters 
$\gamma / (\kappa q_c^4) = 10^{-4}$ and $k/(\kappa q_{c}^{2})=10^{-2},1,10^{2}$ are used. The line types for each $k$ values are given in the panel (b). Competition between the energy functions 
presented in the panel (a) leads to the variety of the fluctuation spectrum as presented in the panel (b), depending on the network rigidity. The inset shows the power law
exponent of the fluctuation spectrum, defined in Eq.~(\ref{alpha}). 
The coupling to extremely soft network ($k=10^{-2} \kappa q_{c}^{2}$) yields insignificant modification of the membrane fluctuation. 
For $k=\kappa q_{c}^{2}$, a smooth crossover from $q^{-4}$ to $q^{-2}$ occurs, as the exponent in the inset displays. For the stiffer network~($k=10^{2}\kappa q_{c}^{2}$)
such crossover is interrupted by higher-order bending energy, resulting in the non-monotonic evolution of $\alpha$.}
\label{fig:SClog}
\end{figure}

{\it Strongly coupled membrane}: If the coupling energy $v$ is sufficiently larger than the cytoskeleton elastic energy, i.e., $v \gg 4k = \mathrm{max}~E_e(q)$, the coupling energy function is given by  
\begin{equation}\label{scc}
C(q) \approx E_e(q)/p^2 \approx kq^{2}, 
\end{equation}
where the last approximation is valid in the long wavelength limit. Figure 1 (a) shows the relevant energy functions for various $k$ values. 
Note that in the region of interest,  among $E_{m}(q+T)$, the term with $T=-2q_{c}$ makes the best contribution to the summation in Eq.~(\ref{C_eff}).
We only consider $E_{m}(q-2q_{c})$ which crosses with $E_{m}(q)$ at the zone boundary $q=q_{c}=\pi/\xi$, as displayed in Fig.~1(a). 
The magnitude of these bending energies in relative to the elastic energy depends not only on $q$ but also on the spring network rigidity $k$,
which according to Eq.~(\ref{cri}) results in the variety of fluctuation spectrum as presented in Fig.~1(b).  We add the inset displaying the power-law exponent $\alpha$,
\begin{equation}
\alpha = - \frac{d \ln \langle |h_{q}|^{2}\rangle}{d\ln q}~,
\label{alpha}
\end{equation} 
in order to see more clearly the wavenumber dependence of the fluctuation spectrum. One common feature among fluctuation spectra for various network rigidities is that in the short wavelength region~($q \gtrsim q_{c}$), the bending energy $E_{m}(q)$ is dominant over $E_{m}(q-2q_{c})$ and $C(q)$, and the fluctuation 
spectrum shows the fluidic behavior $q^{-4}$.

On the other hand, qualitative behavior of the long wavelength fluctuation~($q \lesssim q_{c}$) is crucially altered by the elasticity $k$. 
For small $k/(\kappa q_{c}^{2})=10^{-2}$ (the dashed lines), 
we have that $ E_{m}(q-2q_{c})\gg C(q)$ and $E_{m}(q)\gtrsim C(q)$~(see Fig.~1(a)).  Reminding of the criterion Eq.~(\ref{cri}),
we expect the fluctuation spectrum to be well described by the bare bending energy $E_{m}(q)$, which is indeed so as shown in Fig.~1(b). 
While such soft network does not influence the mechanical properties of the lipid bilayer, for large $k$ there exists a wavenumber
region satisfying $E_{m}(q) < C(q) < E_{m}(q+T)$ (see the curves for $k/(\kappa q_{c}^{2})=1, 10^{2}$ in Fig. 1(a)).
In this case, the elastic energy contribution from the cytoskeleton is observable in the fluctuation spectrum with $q^{-2}$ behavior in Fig.~1(b).

Note that for $k/(\kappa q_{c}^{2}) = 10^{2}$, near the zone boundary $q =q_{c}$, the higher order bending energy $E_{m}(q-2q_{c})$ is comparable or smaller than $C(q)$ and yields the non-monotonic fluctuation spectrum. In a recent paper~\cite{Merath} a fluid membrane discretely coupled to a very hard fluid membrane is found to show non-monotonic fluctuations, and its origin was interpreted by negative surface tension. From our analysis it becomes clear that the non-monotonicity arises from the bending energy for creating short wavelength curvature which is less costly than the deformation
of the extremely rigid cytoskeleton, that is, $E_{m}(q-2q_{c}) < C(q)$. The non monotonic behavior is also reflected on the power law exponent $\alpha$ which has a deep near $q=q_{c}$, while for other $k$ values it monotonically decreases from $\alpha =4$ to $\alpha= 0$ with $\alpha=2$ plateau for $k/(\kappa q_{c}^{2})=1$.

{\it Weakly coupled membrane}: Let us consider that the bilayer-cytoskeleton coupling is not strong enough, i.e., $ v < 4k = \mathrm{max}~ E_e (\mathbf{q})$. Defining a wavenumber $q_{v}$ at which $v=E_{e}(q_{v})$,  we obtain the approximate form of the coupling energy function $C(q)$:
\begin{equation}\label{wcc}
C(q)\approx \left\{ 
\begin{array}{lll}
&E_{e}(q)/p^{2}~,&q \lesssim q_{v} \\
&v/p^{2}~,  &q\gtrsim q_{v}
\end{array} 
\right.
\end{equation}
which well describes the behavior illustrated in Figure 2(a).  This coupling energy should be compared with
the bending energy costs, and the fluctuation nature is again determined by the dominant function which follows Eq.~(\ref{cri}). 

\begin{figure}[!htb]
\resizebox{8.5cm}{!}{\includegraphics{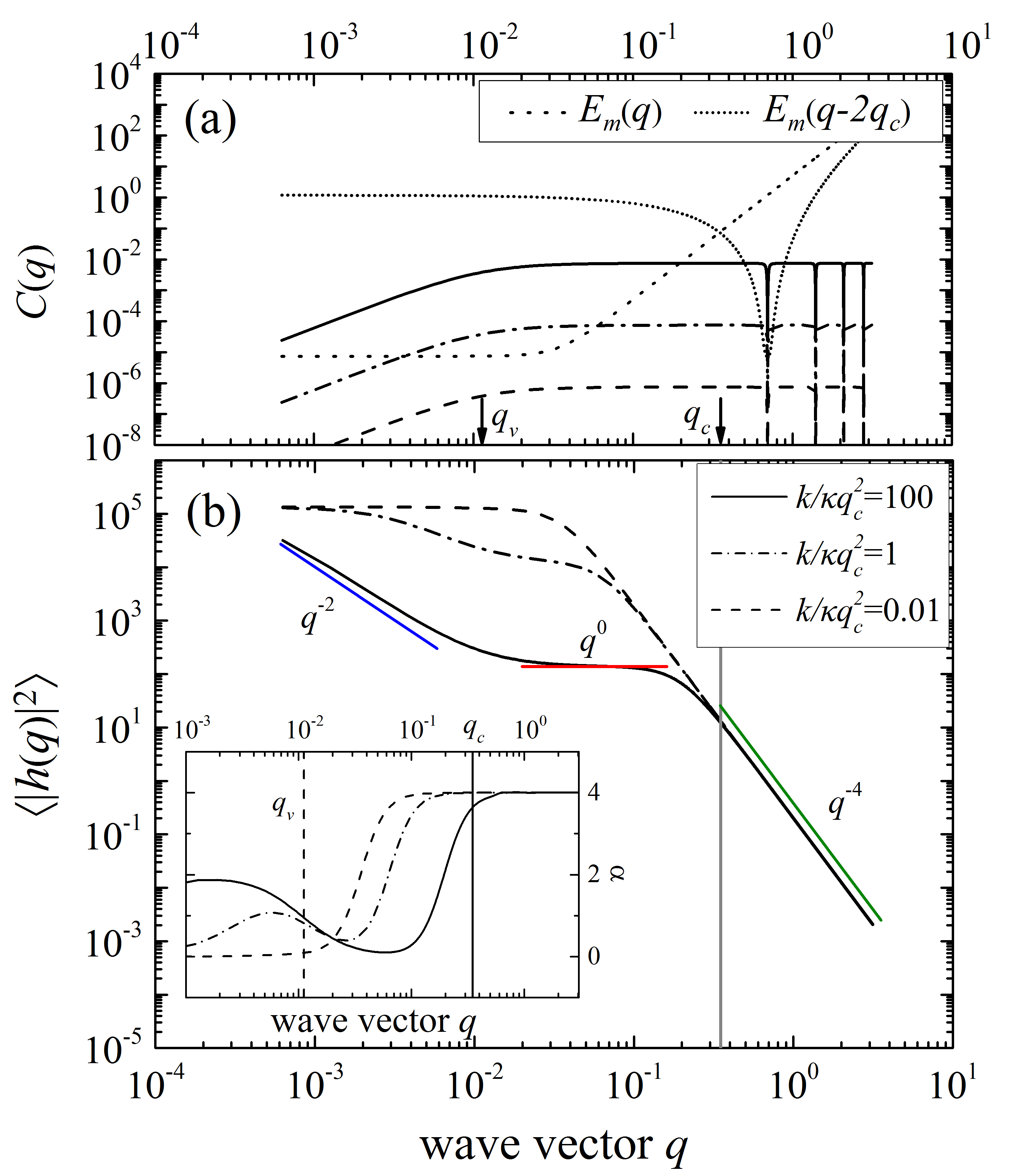}}
\caption{(a) Coupling functions and (b) fluctuation spectra for weakly coupled membranes ($v = 10^{-2}k$), where
the system parameters are same as used in Fig. 1. While for soft networks ($k/(\kappa q_{c}^{2})=10^{-2}, 1$) 
the coupling does not significantly affect the fluctuation properties,  extremely stiff network ($k=10^{2}\kappa q_{c}^{2}$) induces the 
confining effect at the intermediate wavelengths~($q$-independent spectrum) and manifest its elasticity in the fluctuation
spectrum by $q^{-2}$ dependence in the long wavelength region. }
\label{fig:WClog}
\end{figure}

In the short wavelength region~($q \gtrsim q_{c}$), 
the bare bending energy $E_{m}(q)$ is dominant over the other relevant energy functions and results in 
$q^{-4}$ dependent fluctuation spectrum, as displayed in Fig.~2(b). 
In the region $q \lesssim q_{c}$, the
network rigidity comes into play. For small $k$ (see the dashed lines for $k/(\kappa q_{c}^{2}) =10^{-2}$), $E_{m}(q)$ is still
a dominant energy factor, which explains well the fluctuation spectrum presented in Fig.~2(b).
For rather rigid network, at intermediate wavenumber the coupling energy $C(q)$ can be less than $E_{m}(q-2q_{c})$ 
and yet larger than $E_{m}(q)$. In the case, $q$-independent fluctuation can be observable due to the coupling function behavior
for $q \gtrsim q_{v}$ in Eq.~(\ref{wcc}). This coupling induced confining effect is more pronounced for large $k$,  for example, 
$k/(\kappa q_{c}^{2})=10^{2}$. Such rigid network imparts the elasticity for long wavelength ($q \ll q_{v}$) undulation of the membrane 
yielding $q^{-2}$ behavior. This is a unique consequence of the finite coupling strength, which was not expected from the Gov's model \cite{Gov 2003} and other previous studies \cite{dubus}.

\subsection{Confining Potential}

Fluctuation spectrum can also be influenced by the confining potential $\gamma h^2$ in Eq.~(\ref{H_m}). Up to now, although we have considered only very small confining strength $\gamma$ relative to the bending energy $\kappa q^4$, the confining strength of red blood cell membranes is not known precisely.
In this section, we discuss the effect of the confining potential on the fluctuation spectrum. Since the confining potential provides a (positive) constant shift to the bending energy $E_m (q)$, the role of the coupling can be insignificant when the confinement is too strong compared to the cytoskeleton elasticity. In order for
the elastic contribution from the spring network to be visible in the spectrum, in particular, the $q^{-2}$ dependence, the confining should be sufficiently weak,
for which rough estimation gives a condition
\begin{equation}\label{conf}
\sqrt{\gamma/k} \ll  \sqrt{k/\kappa}~(\mbox{for SC}),~ q_{v}~(\mbox{for WC})~.  
\end{equation}
Note here that $\sqrt{\gamma /k}$ is the wavenumber at which the elastic energy $E_{e}(q)$ in the long wavelength limit equals to the confining potential,
and $\sqrt{k/\kappa}$ and $q_{v}$ are roughly the upper boundaries that $E_{e}(q)$ dominates the bending energies for the strong coupling~(SC) and the weak coupling~(WC) case, respectively. 

Different microscopic models have been proposed to understand the physical origin of the cytoskeletal confinement. Inhomogeneous pinning due to sparse connection with the cytoskeleton \cite{Gov 2004} and the closed geometry of RBCs~\cite{Auth, Park model, Nelson} are two major candidates. 
Our results suggest that the two pictures entail very different physical situations.
In our model the discrete coupling,  $v(h-\phi )^2$ in Eq.~(\ref{V}), indeed leads to the $q$-independent fluctuation at long wavelength.
This coupling induced confinement is apparent only if the membrane is weakly coupled to very rigid cytoskeleton (see Fig.~2(b)), and 
in the longer wavelength region~($q < q_{v}$), one should observe the tension signature as well.
For the closed geometry of RBCs, the translational symmetry of the elastic network is broken. This can be effectively described by
adding constant to the elastic energy: $E_{e}(q)\rightarrow \gamma_{0}+E_{e}(q)$, which modifies only the all-to-all coupling kernel $C(q)$.  
Since the geometry induced confining  $\gamma_{0}$ is now part of $C(q)$, for its observability it requires the strong coupling and soft cytoskeleton. 
Such distinctive features depending on the origin of the confinement would help to verify which description is closer to reality.

\subsection{Tension Emergence}

We discuss the tension emergence, comparing our results with previous studies. In Refs.~\cite{Gov 2003, Fournier}, the authors analyzed fluctuation
spectra observed in Ref.~\cite{Zilker}, and focused on the quantity $\kappa/\kappa_{q}$ where $\kappa_{q}$, the effective bending rigidity, is defined as $\langle |h_{\bf q}|^{2}\rangle =k_{B}T/ (\kappa_{q} q^{4})$.  Assuming abrupt increase of surface tension in the long wavelength regime, 
they pointed out that jump occurs in $\kappa/\kappa_{q}$.  Our model does produce neither sudden change of tension nor jump in $\kappa/\kappa_{q}$ curve.  Instead, the tension signature only gradually appears as $q^{-2}$-dependence in $\langle |h_{{\bf q}}|^{2}\rangle$.  
In order to explain the tension jump in our model, the coupling energy function $C(q)$ in Eq.~(\ref{C_q}) should vanish at $q=q_{c}$.
However, in our approach based on the microscopic model, the coupling to the spring network 
causes all-to-all correlations between membrane deformations, as given in Eq.~(\ref{H_eff2}), yielding 
the effective coupling energy $C(q)$ which does not vanish at $q \gtrsim q_{c}$ but becomes less significant than the bare bending energy. 
Therefore, introducing a sudden change of tension at $q_{c}$ in Refs.~\cite{Fournier, Gov 2003} cannot be justified within the physics explained in our model. 

Induced surface tension can also be observed in $\langle |h_{{\bf q}}|^{2}\rangle$ rather than $\kappa/\kappa_{q}$. 
The parameters in Refs.~\cite{Gov 2003, Fournier}, estimated from the fit to the experiment~\cite{Zilker}, is marginal or discordant for observing even the gradual tension emergence in the fluctuation spectrum. When the bending modulus $\kappa = 2 \times 10^{-20} ~\mathrm{J} = 5 ~k_\mathrm{B} T$ \cite{Zilker, Gov 2003, Fournier, Brochard, Park model}, the estimated parameter values are $k /(\kappa q_c^2) \sim O(10^{-2})$ and 
$\gamma/(\kappa q_{c}^{4}) \gtrsim O(10^{-4})$, which gives $\sqrt{\gamma \kappa/k^{2}} \gtrsim O(1) $. 
This does not meet the condition, Eq.~(\ref{conf}), and the signal of induced tension in the fluctuation spectrum should be either weak or absent.
When we use an order of magnitude larger value of the bending modulus, $\kappa = 50 ~k_\mathrm{B} T$, which is also used in recent computer simulations~\cite{kappa 50kbt}, the non-trivial fluctuation spectrum due to the coupling with the cytoskeleton becomes less visible. For a given $k$ and $\gamma$, the effect of different values of $\kappa$ is absorbed to the dimensionless parameters $k/ \kappa q_c^2 $ and $\gamma / \kappa q_c^4$, i.e., increase of $\kappa$ is equivalent to decrease of $k/ \kappa q_c^2 $ and $\gamma / \kappa q_c^4$. In Figure \ref{fig:SClog}, the decrease of $k/\kappa q_c^2$ makes the strongly coupled membrane's peculiar properties like non-monotonic fluctuation near $q=q_c$ and the signature of the induced tension be less visible.  With larger bending modulus, $\sqrt{\gamma \kappa/k^{2}}$ gets greater so that the tension signature becomes more difficult to be observed. Similarly, the Figure \ref{fig:WClog} shows that decrease of $k/\kappa q_c^2$ let the size of the region for wave vector independent fluctuation decrease.
This suggests that, in order to observe the peculiar influences of the cytoskeleton on the fluctuation spectrum, the membrane with soft bending rigidity should be used.

In order to observe $q^{-2}$ behavior, $k$ values should be relatively large. In fact, the spring constant can be estimated in various ways. From the measured shear modulus $\mu\approx 7\times 10^{-6}\mathrm{J/m^{2}}$, 
adopting the continuum model \cite{Park model}, we obtain the spring constant as $k=4\mu/\sqrt{3} \sim 10^{-5} \mathrm{J/m^{2}}$~\cite{bookBoal},
and correspondingly $k/(\kappa q_{c}^{2}) \sim O(1)$. 
If treating the spectrin tetramer as an ideal entropic spring, the spring constant is given by $k_\mathrm{ideal} = 3 k_\mathrm{B} T/2 p L_c $.
The persistence length of the spectrin $p = 7.5 ~\mathrm{nm}$ \cite{Li, Discher 1998} and the contour length 
$L_c = 194 \pm 15 ~\mathrm{nm}$ \cite{Byers} give $k\approx 4 \times 10^{-6}\mathrm{J/m^{2}}$ and $k/(\kappa q_{c}^{2})\sim O(10^{-1})$.
These spring constants are orders of magnitude larger than the fitting values and satisfy Eq.~(\ref{conf}).

\section{Conclusion}
We investigate fluctuation properties of red blood cell membranes through the Gaussian model which takes account of the finite coupling strength between the bilayer and the cytoskeleton. Focusing on the role of the coupling and discrete nature of the spectrin meshwork, we obtain different types of fluctuations for the coupled membrane. A membrane strongly coupled to rigid cytoskeleton presents non-monotonic fluctuations because of the membrane curvature deformation at length scale smaller than the cytoskeleton meshsize. When soft cytoskeleton is coupled to the membrane, the system exhibits the well-studied fluctuation spectrum described by surface tension and the confining potential. Weakly coupled membranes have an extra crossover  and present wave vector independent fluctuation at intermediate length scales. The essential source of this diversity is the competition among the elastic energy of the spring network, the bilayer-cytoskeleton coupling, and the bending energy of the membrane at short length scales.  

Direct measurement of such diverse spectra must be a difficult task, depending on instrument resolution, sample availability, and control technique.
Yet, as shown in this work, the fluctuation spectrum is dependent on the ratio $k / \kappa q_c^2$, and then controlling the bending modulus of red blood cell membranes may allow us to observe non-trivial spectrum like non-monotonic fluctuations for rigid cytoskeleton. ATP concentration can also change the elastic property of the cytoskeleton as well as the coupling strength of the anchoring proteins. More direct evidences should be provided by such controlled experiments for better understanding the physical properties of RBCs.  Moreover, the nonequilibrium nature of RBC flickering
in relation to active coupling between the bilayer and the cytoskeleton is also an important issue. In future studies, we will pursue to analyze nonequilibrium spectrum
in the presence of active coupling, based on our model with a certain type of time dependence introduced in the coupling strength of Eq.~(\ref{V}), which, we believe, is a minimal approach to grasp the essential characteristic of the system dynamics.

\section{acknowledgements}
We thank Y. Park for helpful discussion and providing us unpublished results.
This research was supported by Basic Science Research Program through the National Research Foundation of Korea(NRF) funded by the Ministry 
of Education, Science and Technology(Grant No. NRF-2013R1A1A2013137).


\begin{thebibliography}{10}
\bibitem{note} If the fluctuations are not small as we assumed, there exists a coupling between the in-plane stretching and the bending. Generally the Gaussian approximation is no longer valid for large elastic deformations, and the higher order terms would be necessary.
\end{thebibliography}
\end{document}